\documentclass[twocolumn]{webofc}

\usepackage[varg]{txfonts}   
\usepackage{hyperref}
\usepackage{url}
\usepackage{amsmath}
\usepackage{amssymb}
\usepackage{bm}
\usepackage{siunitx}
\usepackage[usenames,dvipsnames]{xcolor}
\usepackage[symbol*]{footmisc}

\newcommand{\Revised}[1]{{\color{black}#1}}
\newcommand{\Para}[1]{}

\hypersetup{colorlinks=true,citecolor=blue,urlcolor=blue,linkcolor=blue}
\begin{document}
\title{Dislocation Glides in \Revised{Monolayered} Granular Media: Effect of Lattice Constant\thanks{This paper is scheduled to be published in ``EPJ Web of Conferences'' (Powders \& Grains 2025, 8-12 December, Goa, India).}}
%
%

\author{\firstname{Fumiaki} \lastname{Nakai}\inst{1}\fnsep\thanks{\email{fumiaki.nakai@ess.sci.osaka-u.ac.jp}} \and
        \firstname{Takashi} \lastname{Uneyama}\inst{2}\fnsep \and
        \firstname{Yuto} \lastname{Sasaki}\inst{1}\fnsep \and
        \firstname{Kiwamu} \lastname{Yoshii}\inst{3}\fnsep \and
        \firstname{Hiroaki} \lastname{Katsuragi}\inst{1}
}

\institute{Department of Earth and Space Science, Osaka University, 1-1 Machikaneyama, Toyonaka 560-0043, Japan
\and
           Department of Materials Physics, Graduate School of Engineering, Nagoya University, Furo-cho, Chikusa, Nagoya 464-8603, Japan
\and
           Department of Applied Physics, Tokyo University of Science, 6-3-1, Nijuku, Katsushika, Tokyo, 125-8585
          }

\abstract{
A recent study~\cite{Nakai2024-lk} demonstrated that granular crystals containing a single dislocation exhibit dislocation glide analogous to that observed in atomic-scale crystals, resulting in plastic deformation at yield stresses several orders of magnitude lower than those of dislocation-free crystals.
The yielding behavior strongly depends on the interparticle friction coefficient $\mu$: dislocation glide occurs for friction coefficients below a critical value $\mu_c$, while crystalline order deteriorates above $\mu_c$.
In this work, we use discrete element method simulations to systematically investigate how the lattice constant, which determines the interparticle spacing and is a fundamental parameter in microscopic crystalline solids, and the friction coefficient $\mu$ influence the yielding behavior in \Revised{monolayered} granular crystals with dislocation.
By decreasing the lattice constant, we find an increase in the critical friction coefficient $\mu_c$, allowing dislocation glide to persist at higher friction values.
Furthermore, we observe a linear scaling of yield stress with normal stress, except at extremely low friction coefficients.
}
\maketitle
\section{Introduction}
\label{intro}
\begin{figure}
    \centering
    \includegraphics[width=0.8\linewidth]{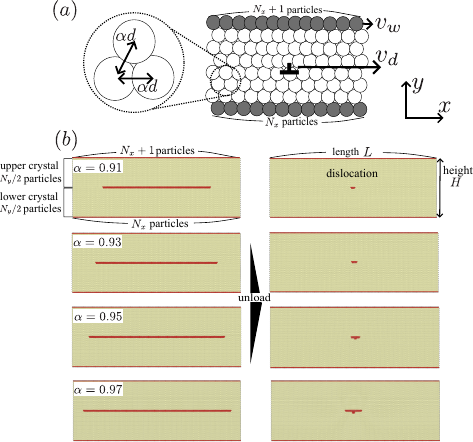}
    \caption{
    (a)
    Schematic illustration of dislocation glide (adapted from Ref.~\cite{Nakai2024-lk}).
    Local deformation propagates similarly to the motion of an inchworm, resulting in a lower yield stress compared to that of a perfect crystal.
    The dislocation core, represented by the symbol $\perp$, moves at a velocity $v_d = N_x v_w$ relative to the wall velocity $v_w$.
    In this study, the distance between neighboring particles is set to approximately $\alpha d$ (see Method section for details), where we define $\alpha$ as the lattice parameter.
    (b) Dislocation core formation in \Revised{monolayered} crystals with varying $\alpha$. The upper and lower crystals contain $(N_x+1)N_y/2$ and $N_x N_y/2$ particles, respectively, with $N_x = 150$ and $N_y = 60$. Yellow particles have six neighbors, while red particles have fewer. Unloading with fixed height $H = \sqrt{3}(N_y+1)d/2$ and width $L = N_x d$ generates the dislocation core, whose width increases with larger $\alpha$. Shear is applied by moving the top wall at velocity $\dot{\gamma} H$.}
    \label{figmain:initial}
\end{figure}

\Para{---1st paragraph---}
In molecular-scale crystals, line defects called dislocations arise due to entropic effects~\cite{anderson2017theory}. When external forces are applied to crystals with dislocations, plastic deformation occurs through dislocation glide—a localized deformation that propagates in an inchworm-like manner. This deformation initiates at yield stresses called Peierls stress, which are several orders of magnitude lower than the theoretical yield stresses of perfect crystals~\cite{anderson2017theory, Peierls1940-ah, Nabarro1947-en, Huntington1955-jc, Joos1997-zn}, as illustrated in Fig.~\ref{figmain:initial}(a).
Plastic deformation in microscopic crystals is governed by dislocation dynamics, an extensively studied phenomenon in materials science.
Our previous work~\cite{Nakai2024-lk} demonstrated that granular media with interparticle friction can exhibit dislocation glide when using a system grounded in the fundamental theories of atomic-scale dislocations~\cite{anderson2017theory, Peierls1940-ah, Nabarro1947-en, Joos1997-zn}.
The system exhibits yield stresses that are significantly lower than those of dislocation-free crystals, analogous to behavior observed in microscopic crystals~\cite{Nakai2024-lk}.
While dislocation glide in granular systems is a novel deformation mode—distinct from both crystalline~\cite{Karuriya2023-me} and amorphous~\cite{Andreotti2013-yo} plasticity—its system parameter dependence remains largely unexplored.

\Para{---2nd paragraph---}
Theoretical models~\cite{Peierls1940-ah, Nabarro1947-en, Joos1997-zn} suggest that the yield stress of microscopic crystalline media is governed by several characteristic length scales: the magnitude of the Burgers vector, the lattice spacing perpendicular to the dislocation line, and the dislocation core width.
These characteristic lengths cannot be independently controlled; instead, they can vary simultaneously when the lattice constant changes.
Specifically, in our previous study~\cite{Nakai2024-lk}, which employed the configuration shown in Fig.~\ref{figmain:initial}(a), these characteristic lengths were determined by the lattice constant—represented by the lattice parameter $\alpha$—provided that the particle material properties remained fixed.
This earlier study, using discrete element method (DEM) simulations, also identified the interparticle friction $\mu$ as another critical factor influencing the deformation of granular crystals with dislocations.
These previous findings raise a fundamental question: How does interparticle friction~$\mu$ influence the yielding behavior of granular systems containing dislocations as the lattice parameter $\alpha$ is systematically varied?

\Para{---3rd paragraph---}
To address this question, we analyze a \Revised{monolayered} system constructed based on fundamental dislocation theories~\cite{anderson2017theory}.
We employ DEM simulation, \Revised{neglecting the gravitational force,} to investigate the structural and rheological responses of the system under shear, exploring a range of interparticle friction coefficients~$\mu$ and lattice parameters~$\alpha$.
\Revised{
In the context of this work, the term "yielding" refers mechanically to the event marked by the first drop in shear stress. This definition encompasses both dislocation glide and crystal breakage.}

\begin{table}[htbp]
\caption{
DEM simulation Parameters (Hertz-Mindlin-Tsuji model) derived from typical elastomers~\cite{callister2020materials}.
\label{table:parameters}}
\begin{tabular}{c|c|c}
\hline
Parameter & Symbol & Value \\ \hline
Density & $\rho$ & \SI{1000}{\kg\per\m^3} \\
Diameter & $d$ & \SI{1}{\mm} \\
Young’s modulus & $E$ & \SI{1}{\MPa} \\
Poisson’s ratio & $\nu$ & 0.45 \\
Restitution coeff. & $e$ & 0.6 \\
Particle number & $N$ & $(2N_x+1)N_y/2$ \\
Box width & $L$ & $N_x \alpha d$ \\
Box height & $H$ & $\sqrt{3}(N_y+1)\alpha d/2$ \\
Shear rate & $\dot{\gamma}$ & \SI{1e-3}{\per\s} \\
Friction coeff. & $\mu$ & $0 \le \mu \le 1$ \\
Lattice parameter & $\alpha$ & $0.91 \le \alpha \le 0.99$ \\
\hline
\end{tabular}
\end{table}

\section{Method}

\Para{---4th paragraph---}
The system configuration follows our previous work~\cite{Nakai2024-lk}, with the simulation methodology detailed below.
We employ the DEM simulation~\cite{Cundall1979-px} as implemented in LAMMPS (Large-scale Atomic/Molecular Massively Parallel Simulator)~\cite{thompson2022lammps}.
Interparticle interactions are modeled by the Hertz-Mindlin-Tsuji model.
The time evolution of the position~$\bm{r}_i$ and angular velocity~$\bm{\omega}_i$ of the $i$-th particle follows Newton’s equations of motion:
\begin{align}
    m \ddot{\bm{r}}_i &= \sum_{i\neq j} \bm{F}_{ij}=
    \sum_{i \neq j} (F_{n,ij}\bm{n}_{ij} + F_{\tau,ij}\bm{\tau}_{ij}) \Theta(d - |\bm{r}_{ij}|) ,\label{eq:motion_translational}\\
    \bm{I} \cdot \dot{\bm{\omega}}_i &=
    \sum_{i \neq j}
    (\bm{l}_{ij}\times \bm{\tau}_{ij})F_{\tau,ij}
    \Theta(d - |\bm{r}_{ij}|).
    \label{eq:motion_rotational}
\end{align}
Here, $d$ denotes the particle diameter, and $m = \rho \pi d^3 / 6$ represents the mass of a spherical particle, where $\rho$ is the mass density.
The inertia tensor~$\bm{I}$ for a uniform sphere is given by $\bm{I} = (m d^2 / 10) \bm{1}$, where $\bm{1}$ is the unit tensor.
The notation~$\dot{X}$ denotes the time derivative of a variable~$X$.
The relative position between the $i$-th and $j$-th particles is defined as $\bm{r}_{ij} = \bm{r}_i - \bm{r}_j$, where $\bm{n}_{ij}$ and $\bm{\tau}_{ij}$ are the unit vectors along the normal and tangential directions, respectively.
The vector~$\bm{l}_{ij}$ extends from the center of the $i$-th particle to its contact point with the $j$-th particle.
The Heaviside step function $\Theta$ activates the contact force.
The normal and tangential forces between the $i$-th and $j$-th particles, $F_{n,ij}$ and $F_{\tau,ij}$, are defined as follows:
\begin{align}
    F_{n,ij} &= k_n\xi^{3/2}_{n,ij}
    + \eta v_{n,ij}, \label{eq:force_normal}\\
    F_{\tau,ij} &= 
    \min\left(\left|k_{\tau} \sqrt{\xi_{n,ij}}
    \bm{\xi}_{\tau,ij} + \eta \bm{v}_{\tau,ij}\right|, 
    \mu  F_{n,ij}\right),
    \label{eq:force_tangential}
\end{align}
where the elastic constants along the normal and tangential directions are given by $k_n = \frac{E \sqrt{d}}{3 (1 - \nu^2)}$ and $k_\tau = \frac{E \sqrt{d}}{(1 + \nu)(2 - \nu)}$, respectively, where $E$ and $\nu$ denote Young’s modulus and Poisson’s ratio, respectively.
The relative velocities along the normal and tangential directions at the contact point between the $i$-th and $j$-th particles are defined as $v_{n,ij} = -\bm{v}_{ij} \cdot \bm{n}_{ij}$ and $\bm{v}_{\tau,ij} = \bm{v}_{ij} - (\bm{v}_{ij} \cdot \bm{n}_{ij}) \bm{n}_{ij} - \frac{d}{2} (\bm{\omega}_i + \bm{\omega}_j) \times \bm{n}_{ij}$, respectively.
The damping constant~$\eta$ is set to the same value for both normal and tangential forces, consistent with previous studies~\cite{Marshall2009-yp}.
We define $\eta=\kappa \sqrt{m k_n\sqrt{\xi_{n,ij}}/2}$, where $\kappa=1.2728-4.2783e+11.087e^2-22.348e^3+27.467e^4-18.022e^5+4.8218e^6$ ensures a constant restitution coefficient $e$.
The quantities $\xi_{n,ij}$ and $\bm{\xi}_{\tau,ij}$ represent the normal overlap and accumulated tangential displacement, respectively~\cite{luding2008cohesive}.
The $\min$ function enforces Coulomb’s friction criterion using the interparticle friction coefficient~$\mu$.
The discretized timestep size is set to $\SI{8.9e-6}{s}$.

\Para{---5th paragraph---}
Following our previous work~\cite{Nakai2024-lk}, we construct an initial configuration featuring a \Revised{monolayered} hexagonal lattice with a dislocation.
As shown in Fig.~\ref{figmain:initial}(b), the lower hexagonal crystal, comprising $N_x N_y / 2$ bulk particles, is aligned with lattice vectors $(\alpha d, 0)$ and $(\alpha d / 2, \sqrt{3} \alpha d / 2)$, while the upper crystal, with $(N_x + 1) N_y / 2$ bulk particles, follows $(\alpha d N_x / (N_x + 1), 0)$ and $(\alpha d N_x / [2 (N_x + 1)], \sqrt{3} \alpha d N_x / [2 (N_x + 1)])$.
The top and bottom particles, which share the same material properties as the bulk particles, act as wall particles.
\Revised{
To prepare the initial configuration, the bottom wall particles are fixed in both the $x$- and $y$-directions. The top wall particles are fixed in the $y$-direction (thus fixed height $H$) and are constrained to move together along the $x$-axis as a single unit.
Allowing the system to relax according to the equations of motion Eqs~\eqref{eq:motion_translational} and \eqref{eq:motion_rotational}, the misalignment over the interface concentrates on the center, and a single dislocation is obtained (see Fig.~\ref{figmain:initial}(b)), resulting in zero apparent shear stress at the walls.
In this relaxation procedure, we wait $\SI{89}{s}$ ($10^7$ time-steps).
}
\Revised{In Supplemental material, the changes in normal stress $\sigma_{yy}$ during the relaxation process are presented with various $\alpha$ in Fig.S1. The relaxation times for each $\alpha$ are significantly shorter than the waiting time ($\SI{89}{s}$).
}
The dislocation core width increases with $\alpha$, as illustrated in Fig.~\ref{figmain:initial}(b), which is qualitatively consistent with the theoretical model~\cite{anderson2017theory}.
From these initial configurations, the wall particles are displaced at a velocity~$v_w = \dot{\gamma} H$ to impose a shear rate of $\dot{\gamma} = \SI{1e-3}{\per\second}$ \Revised{with the fixed height $H$}.
We investigate the structural and rheological responses across a range of $\alpha$ and $\mu$.

\section{Results and Discussion}

\begin{figure}[htbp]
    \centering
    \includegraphics[width=0.8\linewidth]{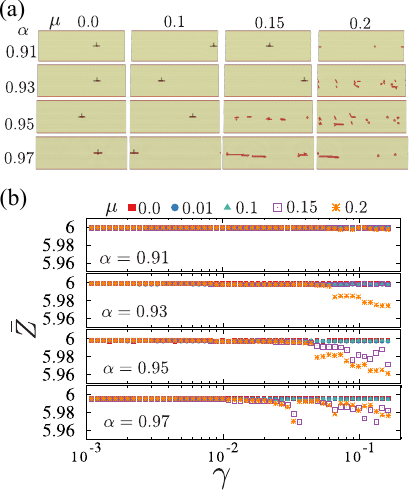}
    \caption{
    (a) System structures after a shear strain of $\gamma \simeq 0.13$ for varying interparticle friction $\mu$ and lattice parameter $\alpha$.
    The dislocation core, represented by the symbol $\perp$, glides along the $x$-axis for small $\mu$ and $\alpha$, whereas glide does not occur for large $\mu$ or $\alpha$ due to crystal fracture.
    \Revised{
    Animations corresponding to these simulations are available in the Supplemental Material.}
    (b) Mean coordination number $\bar{Z}$ versus shear strain $\gamma$ for varying $\mu$ and $\alpha$. When dislocation glide maintains a stable hexagonal structure, $\bar{Z}$ remains close to 6. As crystal order breaks, as shown in Fig.~\ref{figmain:structures}(a), $\bar{Z}$ deviates from 6 with increasing $\mu$ or $\alpha$.}
    \label{figmain:structures}
\end{figure}

\Para{---6th paragraph---}
The \Revised{structural response} varies significantly with both the lattice parameter~$\alpha$ and the interparticle friction~$\mu$, as illustrated by snapshots at a strain of $\gamma \simeq 0.13$ in Fig.~\ref{figmain:structures}(a)\Revised{; corresponding animations are available in the SM}.
Dislocation glide occurs at low values of $\mu$ and $\alpha$.
At high values of $\mu$, dislocation glide is suppressed due to the breakdown of the crystalline structure.
The critical value of $\mu_c$ at which the transition from dislocation glide to crystal breakage occurs decreases with increasing $\alpha$, making dislocation glide less likely at larger $\alpha$.
This behavior is quantitatively captured by the mean coordination number~$\bar{Z}$ as a function of strain~$\gamma$, as shown in Fig.~\ref{figmain:structures}(b).
For $\alpha = 0.91$, $\bar{Z}$ remains approximately $6$ across all tested values of $\mu$, except at $\mu = 0.2$, where $\bar{Z}$ exhibits slight fluctuations around $6$ due to disruptions in crystal symmetry, as depicted in the snapshot for $\alpha = 0.91$ and $\mu = 0.2$ in Fig.~\ref{figmain:structures}(a).
As $\alpha$ increases, the threshold value of $\mu$\Revised{,} at which $\bar{Z}$ deviates from $6$\Revised{,} decreases, indicating that the critical $\mu_c$\Revised{,} marking the transition from dislocation glide to crystal breakage\Revised{,} depends on $\alpha$.
\Revised{The detailed determination of $\mu_c$ as a function of $\alpha$ is left for future investigation.}

\begin{figure}
    \centering
    \includegraphics[width=0.75\linewidth]{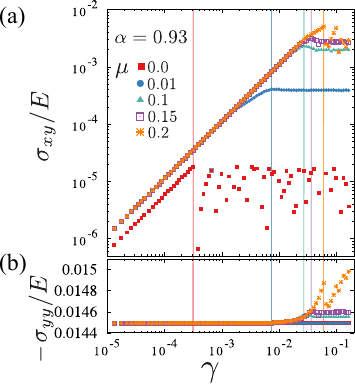}
    \caption{
    (a) Shear stress normalized by Young's modulus, $\sigma_{xy}/E$, versus strain $\gamma$ for various interparticle friction coefficients $\mu$ at a fixed lattice parameter $\alpha=0.93$.
    Vertical lines indicate the strains at which initial stress drops occur.
    The corresponding shear stress, $\sigma_{xy}^*$ shown in Fig.~\ref{figmain:yield}(a), represents the yield stress, which decreases with smaller $\mu$.
    (b) Normal stress normalized by Young's modulus, $\sigma_{yy}/E$, versus strain $\gamma$. In this constant-volume setup, an increase in $\sigma_{yy}/E$ with increasing $\gamma$ is observed. For sufficiently small $\mu$, no increase in $\sigma_{yy}/E$ occurs, as dislocation glide relaxes the system before increasing $\sigma_{yy}/E$. At intermediate $\mu$ (0.1 or 0.15), a slight increase in $\sigma_{yy}/E$ is observed despite the presence of dislocation glide (see Fig.~\ref{figmain:structures}(a)), while at higher friction $\mu \simeq 0.2$, significant increase in $\sigma_{yy}/E$ reflects the breakdown of crystalline order.
    }
    \label{figmain:stress}
\end{figure}

\Para{---7th paragraph---}
To characterize the rheology of the current system, Fig.\ref{figmain:stress}(a) presents the shear stress scaled by the Young’s modulus, $\sigma_{xy}/E$, as a function of strain $\gamma$ for various $\mu$ at $\alpha = 0.93$.
The quantity $\sigma_{xy}/E$ initially exhibits a linear dependence on $\gamma$, followed by a drop at specific strain values marked by vertical lines; the corresponding shear stress at these points defines the yield stress, denoted as $\sigma^{*}_{xy}/E$.
As the interparticle friction coefficient~$\mu$ decreases, the yield stress~$\sigma^{*}_{xy}/E$ also decreases, and this trend is consistently observed across various values of $\alpha$ (data not shown for clarity).
Fig.~\ref{figmain:stress}(b) displays the scaled normal stress, $\sigma_{yy}/E$, as a function of strain~$\gamma$.
For sufficiently small values of $\mu$, an increase in $\sigma_{yy}/E$ does not occur, as deformation relaxes through dislocation glide before increasing $\sigma_{yy}/E$.
Meanwhile, a slight increase in $\sigma_{yy}/E$ is observed for intermediate friction coefficients ($\mu = 0.1$ and $0.15$), corresponding to the regime of dislocation glide, as shown in Fig.~\ref{figmain:structures}(a).
At sufficiently large values of $\mu$, crystal failure occurs, leading to a significant increase in the scaled normal stress~$\sigma_{yy}/E$.

\begin{figure}
    \centering
    \includegraphics[width=0.8\linewidth]{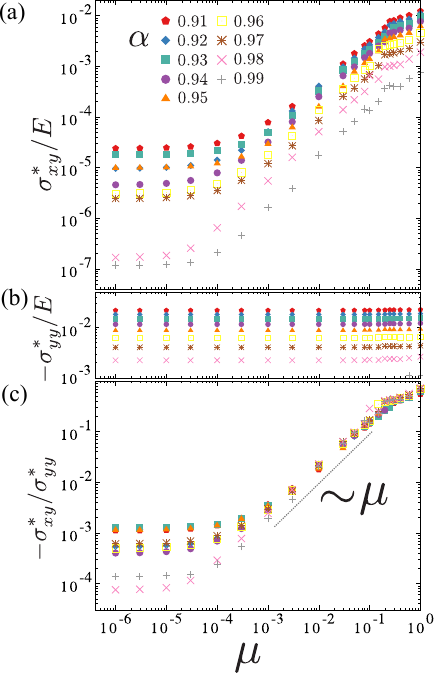}
    \caption{
    (a) Scaled yield stress $\sigma^{*}_{xy}/E$ as a function of interparticle friction $\mu$ for various lattice parameters $\alpha$.
    $\sigma^{*}_{xy}/E$ remains constant at small $\mu$ and exhibits a linear relationship with $\mu$ at intermediate values, consistent with previous findings, across computed $\alpha$.
    In the linear regime, $\sigma^{*}_{xy}/E$ increases monotonically with increasing $\alpha$, whereas in the constant regime, $\sigma^{*}_{xy}/E$ exhibits a nonmonotonic dependence on $\alpha$.
    (b) Scaled normal stress at the first drop in shear stress, $\sigma^{*}_{yy}/E$ (see Fig.~\ref{figmain:stress}(b)).
    $\sigma^{*}_{yy}/E$ increases slightly with increasing $\mu$ for all $\alpha$ and increases monotonically with decreasing $\alpha$.
    (c) Shear stress normalized by normal stress, $\sigma^{*}_{xy}/\sigma^{*}_{yy}$, as a function of $\mu$. In the linear regime, $\sigma^{*}_{xy}/\sigma^{*}_{yy}$ scales consistently, whereas in the constant regime, it does not scale.
    }
    \label{figmain:yield}
\end{figure}

\Para{---8th paragraph---}
Using data from Figs.~\ref{figmain:stress}(a,b), the shear and normal stresses at the yielding point, $\sigma^{*}_{xy}/E$ and $\sigma^{*}_{yy}/E$, are plotted as functions of the interparticle friction coefficient~$\mu$ for various lattice parameters~$\alpha$ in Figs.~\ref{figmain:yield}(a,b).
The yield stress~$\sigma^{*}_{xy}/E$ remains constant for sufficiently small $\mu$ and exhibits a linear dependence on $\mu$ in the intermediate range.
This behavior is consistently observed across all examined lattice parameters~$\alpha$.
In the linear regime, $\sigma^{*}_{xy}/E$ decreases monotonically with increasing $\alpha$, whereas in the constant regime at small $\mu$, its dependence on $\alpha$ is nonmonotonic.
The scaled normal stress $\sigma^{*}_{yy}/E$ monotonically increases as the lattice parameter~$\alpha$ decreases, reflecting the enhanced stiffness associated with denser particle arrangements. Although the dependence of $\sigma^{*}_{yy}/E$ on the friction coefficient~$\mu$ is finite (see also Fig.~\ref{figmain:stress}(b)), it remains minor.
Figure~\ref{figmain:yield}(c) presents the shear stress normalized by the normal stress at yielding, $\sigma^{*}_{xy}/\sigma^{*}_{yy}$, as a function of the friction coefficient~$\mu$ for various $\alpha$.
In the linear regime at intermediate $\mu$, the normalized shear stress scales as $\sigma^{*}_{xy}/\sigma^{*}_{yy} \simeq \mu$, consistent with a model from our previous work~\cite{Nakai2024-lk}.
However, in the constant regime at small $\mu$, no simple scaling relation applies, and the dependence of $\sigma^{*}_{xy}/\sigma^{*}_{yy}$ on $\alpha$ is nonmonotonic.

\Para{---9th paragraph---}
At intermediate values of the friction coefficient~$\mu$, the yield stress scales linearly with the normal stress and is proportional to~$\mu$, indicating that interparticle friction is the dominant factor controlling the yielding behavior.
Such behavior is characteristic of dislocation glide observed in granular crystals; amorphous granular materials do not exhibit this clear proportionality to interparticle friction~\cite{Da_Cruz2005-lr}, as elastic contributions become non-negligible compared to frictional effects.
In contrast, at very small $\mu$, the yielding behavior is dominated by the elastic barrier associated with the Peierls stress, rather than by interparticle friction.
The magnitude of the elastic barrier contribution is sensitive to details of the system, as demonstrated by its dependence on the lattice parameter~$\alpha$ in Fig.~\ref{figmain:yield}(c).
Although not investigated in this study, variations in the functional form of the interparticle potential or in material-specific parameters could further influence the magnitude of this elastic barrier.

\section{Conclusion}
\Para{---10th paragraph---}
In this study, using the DEM simulation, we investigated the effect of the lattice parameter~$\alpha$ on the yielding behavior of a \Revised{monolayered} granular crystal containing a dislocation across various interparticle friction~$\mu$.
Dislocation glide occurs at small $\mu$, whereas it is suppressed when $\mu$ exceeds a critical friction coefficient $\mu_c$.
We found that the critical friction $\mu_c$ decreases with increasing lattice parameter~$\alpha$, suggesting that dislocation glide is favored in denser configurations.
Moreover, by varying $\alpha$, we demonstrated that, during dislocation glide, the yield stress scales linearly with the normal stress in the intermediate friction regime.
However, this scaling relation breaks down in the constant-yield-stress regime at very small $\mu$, where the elastic barrier dominates.
These findings could inform experimental design and control of dislocation glide phenomena in granular media.

\bibliography{ref}

\end{document}